\documentclass[twocolumn]{aastex701}

\usepackage{graphicx} 
\usepackage[para]{threeparttable}
\usepackage{booktabs}

\begin{document}


\title{A VLA Study of the Disturbed Massive Cluster PLCK~G165.7\,+\,67.0 and Its Two Narrow Angle Tail Galaxies }

\author[0009-0002-4145-1011]{Aurora Wilde}
\affiliation{Steward  Observatory, University of Arizona 933 N Cherry Ave Tucson, AZ 85719 USA}
\affiliation{Department of Physics, Massachusetts Institute of Technology, 77 Massachusetts Avenue, Cambridge, MA 02139, USA}
\email{aurora.elle.wilde@gmail.com}

\author[0000-0002-2640-5917]{Eric F. Jim\'{e}nez-Andrade}
\affiliation{Instituto de Radioastronomía y Astrofísica, Universidad Nacional Autónoma de México, Antigua Carretera a Pátzcuaro \# 8701, Ex-Hda. San José de la Huerta, Morelia, Michoacán, México C.P. 58089, Mexico}
\email{e.jimenez@irya.unam.mx}

\author[0000-0003-1625-8009]{Brenda L. Frye}
\affiliation{Steward Observatory, University of Arizona 933 N Cherry Ave Tucson, AZ 85719 USA}
\email{bfrye@arizona.edu}

\author[0000-0001-9394-6732]{Patrick S. Kamieneski}
\affiliation{School of Earth and Space Exploration, Arizona State University, Tempe, AZ 85287-6004, USA}
\affiliation{Department of Space, Earth \& Environment, Chalmers University of Technology, SE-412 96 Gothenburg, Sweden}
\email{pkamiene@asu.edu}

\author[0000-0001-5429-5762]{Kevin C. Harrington}
\affiliation{Joint ALMA Observatory, Alonso de C\'{o}rdova 3107, Vitacura, Casilla 19001, Santiago de Chile, Chile}
\email{kcorneil1223@gmail.com}

\author[0000-0002-2808-0853]{Megan Donahue}
\affiliation{Michigan State University, Department of Physics \& Astronomy, 567 Wilson Road, East Lansing, MI 48824, USA}
\email{donahu42@msu.edu}

\author[0000-0001-7095-7543]{Min S. Yun}
\affiliation{Department of Astronomy, University of Massachusetts, Amherst, MA 01003, USA}
\email{myun@umass.edu}

\author[0000-0002-6610-2048]{Anton M. Koekemoer}
\affiliation{Space Telescope Science Institute, 3700 San Martin Drive, Baltimore, MD 21218, USA }
\email{koekemoer@stsci.edu}

\author[0000-0002-9838-8191]{Marcel Neeleman}
\affiliation{National Radio Astronomy Observatory, 520 Edgemont Rd, Charlottesville, VA 22903 USA}
\email{mneelema@nrao.edu}

\author[0000-0002-2282-8795]{Massimo Pascale}
\affiliation{Department of Astronomy, University of California, 501 Campbell Hall \#3411, Berkeley, CA 94720, USA }
\email{massimopascale@berkeley.edu}

\author[0000-0002-7460-8460]{Nicholas Foo}
\affiliation{School of Earth \& Space Exploration, Arizona State University, Tempe, AZ 85287-1404, USA}
\affiliation{Department of Astronomy/Steward Observatory, University of Arizona, 933 N. Cherry Avenue, Tucson, AZ 85721, USA}
\email{nickfoo@arizona.edu}

\author[0009-0001-7446-2350]{Reagen Leimbach}
\affiliation{Department of Astronomy \& Astrophysics, University of California San Diego, 9500 Gilman Drive La Jolla, CA 92093 USA}
\email{rleimbachmurray@ucsd.edu}

\author[0000-0003-3329-1337]{Seth Cohen}
\affiliation{School of Earth and Space Exploration, Arizona State University, Tempe, AZ 85287-6004, USA}
\email{seth.cohen@asu.edu}

\author[0000-0001-8156-6281]{Rogier A. Windhorst}
\affiliation{School of Earth \& Space Exploration, Arizona State University, Tempe, AZ 85287-1404, USA}
\email{windhors@asu.edu}

\begin{abstract}

Since spectroscopic measurements provide only the line-of-sight velocity, transverse motions are difficult to constrain. In the vast majority of cases, transverse velocities have been measured only for local galaxies. Radio galaxies whose jets are bent back by ram pressure into nearly parallel Narrow-Angle Tails (NATs) offer a way to extend such measurements to higher redshifts. Here, we present proprietary 15~GHz (Ku-band) and archival 6~GHz (C-band) continuum observations from the Very Large Array (VLA) of six radio galaxies in the galaxy cluster field PLCK G165.7+67.0 (G165). VLA imaging resolves two of the cluster galaxies into NATs whose tails both extend toward the Southwest, away from the cluster center of mass. Spectral index maps made at matched angular resolution resolve a steepening gradient consistent with radiative aging along the length of the tails. Physical properties of the NATs are measured and used to constrain two independent models: a Mach cone model and a nonrelativistic hydrodynamic flow model based on Euler's equation. The models returned transverse velocities of $\sim 2000 ~\rm km~ \rm s^{-1}$ for both galaxies. One NAT is the brightest cluster galaxy (BCG) and has a measured radial velocity of $-3300 ~\rm km ~\rm s^{-1}$, resulting in an even higher space velocity. The high inferred 3D velocity of the BCG may be explained if it is intercepted close to core passage.

\end{abstract}

\section{Introduction}

\begin{figure*}
    \centering
    \includegraphics[width=\textwidth]{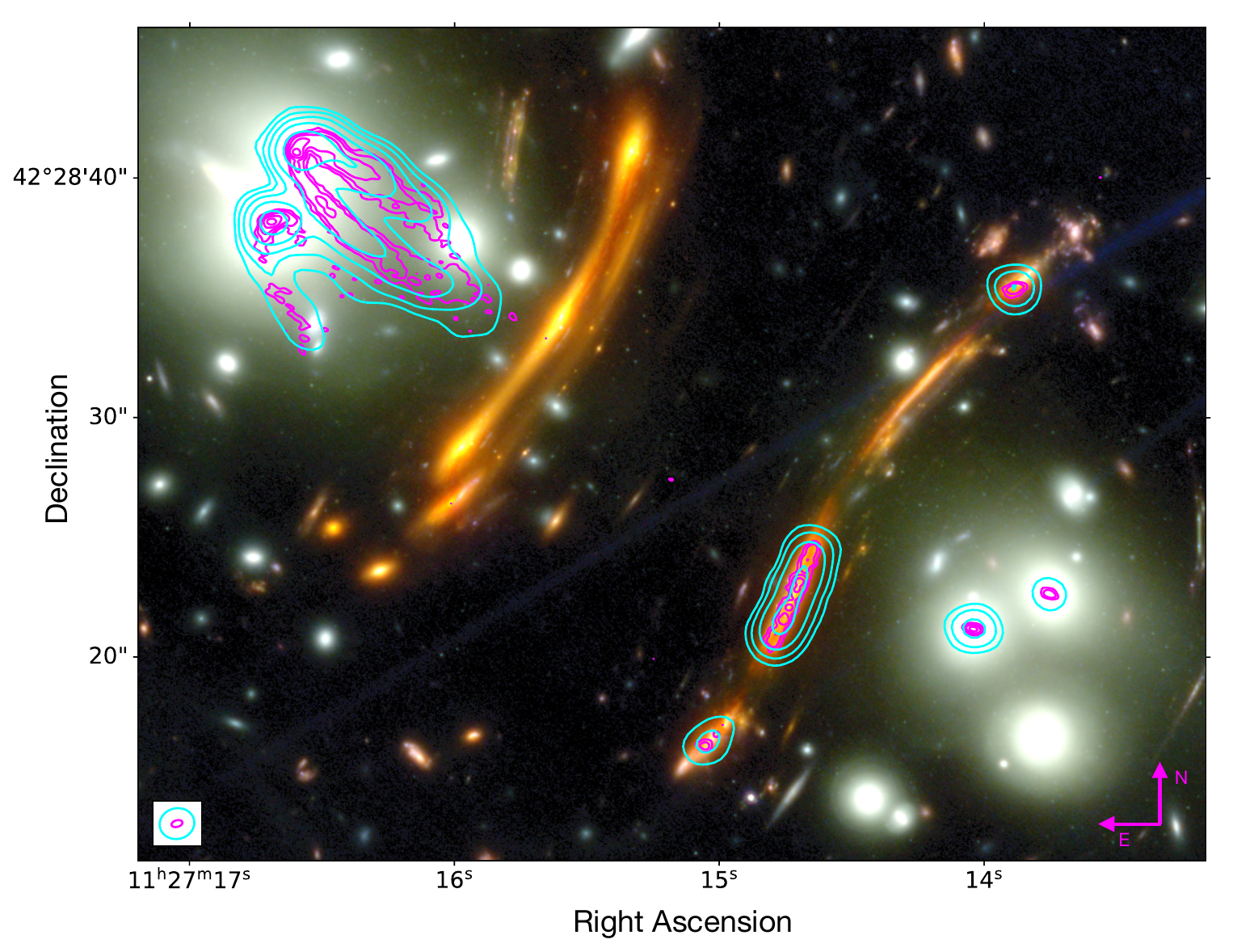}
    \caption{North-aligned JWST NIRCam five-epoch color image of G165 with (R, G, B) = (F444W + F410M+ F356W, F277W + F200W, F150W + F115W+ F090W). VLA C-band contours and corresponding beam are shown in pink, with the contours start from $3\sigma$ and increase by factors of 2. Similarly, Ku-band contours and beam are shown in teal with contours starting at $3\sigma$ and increasing by factors of 2.    }
    \label{fig:nircam}
\end{figure*}

Three-dimensional space velocities are a challenge to obtain for extragalactic sources since Doppler shifts constrain only line-of-sight motion, leaving the transverse velocity component ($v_{gal,t}$) unknown. For the Andromeda galaxy, multi-epoch astrometry using the Hubble Space Telescope (HST) has enabled direct proper-motion measurements, determining $v_{gal,t} \sim 40\text{--}100\text{ km s}^{-1}$ \citep{Sohn_2012, vanderMarel_2012}. Similar techniques have also constrained transverse velocity for M33 \citep{Brunthaler_2005}, the Magellanic Clouds \citep{Kallivayalil_2006}, and local dwarf spheroidals \citep{Gaia_2018}. Direct astrometric techniques, however, become impractical at cosmological distances, where angular displacements are so small that they fall below the detection limits. If a distant galaxy can leave a detectable imprint on the sky that was caused by a dynamical process, then in principle a transverse velocity can be obtained.

Narrow Angle Tail (NAT) galaxies are a distinct morphological class of radio galaxies characterized by a pair of radio jets which are bent backward into a nearly parallel configuration (opening angle $< 90^\circ$) \citep{O'Dea_2023}. NATs are typically associated with elliptical galaxies, have almost exclusively been observed in galaxy groups or clusters, and are thought to require dynamic environments (e.g. disturbances by mergers) to form \citep{Jaffe_1973, Begelman_1979, Bliton_1998, O'Dea_2023, VanDerJagt_2025}. The frequent detection of X-ray substructure in clusters hosting NATs indicates a strong correlation between active cluster mergers and NAT formation \citep{Bliton_1998}. The morphology of the NAT is determined by the galaxy's high peculiar velocity with respect to the 
intracluster medium (ICM), which exerts a ram pressure to sweep the radio jets back behind the galaxy \citep{Jones_1979, O'Dea_1986, O'Neill_2019, O'Dea_2023}. This interaction between the radio jets and their surrounding medium is influenced by the galaxy's velocity, the density and temperature of the ICM, the power and orientation of the jets, and the evolutionary state of the radio source \citep{Douglass_2008, Hu_2021, Rajpurohit_2024_abell}. NAT galaxies thus serve as valuable probes of both the dynamics of galaxy clusters and the properties of the intracluster medium \citep[i.e.,][]{Gani_2026}.

Early models of NATs were derived by assuming equilibrium gas flow between nonrelativistic plasma ejected by the galaxy and the ICM \citep{Miley_1972, Begelman_1979}. The morphology was approximated by a Mach cone \citep{Jaffe_1973, Begelman_1979}, an approach that was later adapted to allow an inference of the transverse velocity \citep{Schellenberger_2017, Rajpurohit_2024}. While the jets were initially treated as blobs of plasma, the presence of magnetic fields in the tails led authors to consider the jet behavior as a fluid \citep{Jaffe_1973, ODea_1985}. Consequently, dynamic and buoyant forces can be assumed in equilibrium and Euler's equation can be applied to model NAT morphology \citep{Burns_1982, ODea_1985}. Others have used measurements of the magnetic field in NATs to measure their lifetimes and infer a velocity from their size \citep{Miley_1980}. Theoretical works predicted that NAT formation necessitates high transverse galaxy velocities, $1-3 \times10^3 ~\rm km ~\rm s^{-1} $ \citep{Begelman_1979, ODea_1985, Bliton_1998}.

In addition to probing cluster kinematics, the spectral evolution of NAT tails can be used to trace the momentum exchange between jet and ICM. As the tails propagate away from the central engine, they undergo radiative losses reflected in a steepening of the radio spectrum \citep{Feretti_1999, Rajpurohit_2024, Gani_2026}. Departures from monotonic aging indicate a secondary injection of energy into the electron population from the ICM, referred to as re-acceleration \citep{VanWeeren_2017}. Mapping spectral index and identifying locations of re-acceleration (or the lack thereof) provides a method for characterizing the nature of structures in the ICM like shock fronts and turbulence \citep{VanWeeren_2017, Rajpurohit_2024}.

Many observational studies have utilized these techniques to measure NAT transverse velocities and characterize the clusters they reside in. For instance, models derived from Euler's equation have been applied to WATs and NATs to infer a transverse velocity \citep{O'Dea_1986, Venkatesan_1994, Sakelliou_1999, Douglass_2008, Hu_2021}. Mach numbers have also been measured for a variety of sources, including the NAT in NGC 741 and 3C 129, providing velocity measurements \citep{Murgia_2016, Schellenberger_2017, Rajpurohit_2024}. In some cases, magnetic field measurements have provided a way to measure the lifetimes of these sources, and in combination with their Mach numbers, a transverse velocity \citep{Murgia_2016, Sebastian_2017, Bruno_2024}. Of these observational studies, many have measured $ v_{\rm gal,t} > 1000 ~\rm km ~\rm s^{-1}$ as expected from theoretical works, though others have inferred smaller velocities of $\sim 600 ~\rm km ~\rm s^{-1}$\citep{O'Dea_1986, Venkatesan_1994, Douglass_2008}. Some studies have used spectral index measurements to identify locations of reaccelerated electrons in NAT tails and radio relics \citep{Murgia_2016, Feretti_1998, Loi_2017, Loi_2020}. On statistical scales, tailed galaxies can be also be used to characterize cluster environments more broadly, including intra-cluster medium density, magnetic field structure, and dynamical state \citep{Johnston-Hollitt_2015}.

Here we apply these diagnostics — transverse velocity measurements and spectral index mapping — to the massive lensing galaxy cluster PLCK G165.7$+$67.0 (G165) ($z=0.348 \pm 0.018$; \citealt{Pascale_2022}), which  hosts two NATs in projected proximity. The cluster is unusual for strongly-lensing an infrared galaxy at $z=2.24$ that was discovered by its rest-frame far infrared colors \citep{Planck_2016, Planck_2020},  confirmed by Herschel Space Observatory \citealt{Canameras_2015, Harrington_2016}), and followed up also by the Hubble Space Telescope (HST) and James Webb Space Telescope (JWST)  Near Infrared Camera (NIRCam) observations \citep{Frye_2019, Pascale_2022, Frye_2024}.  G165 is a disturbed cluster, composed of two distinct mass components. \citet{Frye_2019} cite an upper limit on the radial velocity difference between these components to be $2000 ~\rm km ~\rm s^{-1}$ (see their Figure 10) and in a more detailed spectroscopic follow-up, no discernible radial velocity offset has between the components, suggesting that the cluster-cluster interaction is oriented in the plane of the sky \citep{Pascale_2022, Frye_2024}. Karl G. Jansky Very Large Array (VLA) observations of the cluster uncovered radio-loud galaxies across both components: two radio (NAT) galaxies in the northeast component, and two radio galaxies in the southwest component.  The radio contours for all four galaxies are shown superimposed onto the JWST NIRCam image in Figure \ref{fig:nircam}. While the low X-ray luminosity and Sunyaev–Zeldovich (SZ) decrement of this cluster field may suggest a pre-merger or early infall phase \citep{Frye_2019}, the presence of two NAT galaxies, which are typically formed in high-velocity post-shock environments, suggests a more complex or advanced dynamical history \citep{Pascale_2022}. Analysis of these NATs can provide insights into the overall cluster dynamics and assembly history.


In this work, we analyze VLA data from both C- and Ku-band (6 GHz \& 15 GHz) for the G165 cluster. Our objective is to constrain the transverse velocity of both NATs in the northeast component of the cluster, using both a Mach cone model and a nonrelativistic hydrodynamic flow model. A spectral index map is produced to analyze the spectral aging of the tails and search for evidence of cluster shocks. With this analysis, we hope to better constrain the dynamic environment around the NATs. This paper is organized as follows: Section \ref{sec: data redux} details the data reduction and imaging processes. In Section \ref{sec: results}, we provide measurements and analysis of our images, which are applied to compute transverse velocities. Discussion of the merging state and dynamics of the cluster is found in Section \ref{sec:discussion}, and Section \ref{sec: conc} provides the summary and conclusion. We use a flat $\Lambda$CDM cosmology with $H_0 = 67 ~\rm km~ \rm s^{-1}~ \rm Mpc^{-1}$, $\Omega_m = 0.32$. This corresponds to about 5 kpc per $1^{\prime\prime}$ at $z=0.348$ (cluster member redshift).

\section{Observations} \label{sec: data redux}

\begin{figure*}
    \centering
    \includegraphics[width=\textwidth]{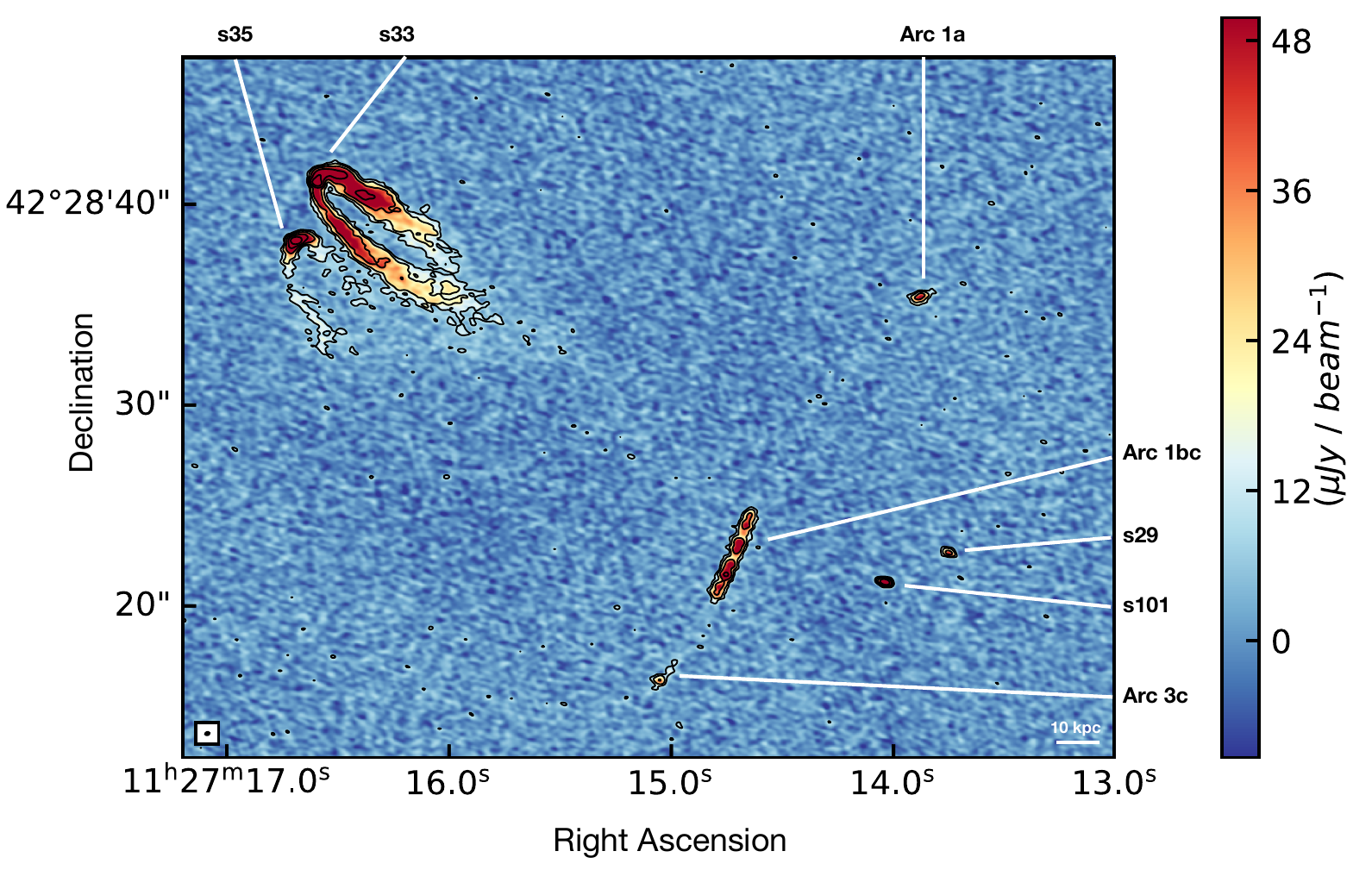}
    \caption{High resolution C-band VLA image of the central region of the G165 cluster, detecting four radio galaxies (s33, s35, s101, and s29), and three images of two galaxies behind the cluster (Arc 1a, Arc 1bc, and Arc 3c). s35 is the BCG of the cluster. The contours shown start at $3\sigma$ and increase by factors of 2. Beam is in the lower left corner. This is the same field of view and orientation as Figure \ref{fig:nircam}. The scale bar in the lower right assumes $z=0.348$.}
    \label{fig:cbnd}
\end{figure*}

\subsection{VLA Data Reduction}

We analyze proprietary and archival VLA observations of the G165 cluster at 6~GHz (C-band) and 15~GHz (Ku-band). We present new 15~GHz observations and the 6~GHz C-configuration data, together with a new reduction of the 6~GHz A-configuration observations previously presented by \citet{Pascale_2022} and \citet{Kamieneski_2024}. The A-configuration data were re-reduced to maximize the signal-to-noise ratio and recover the full extent of the radio tails. All data were processed using the VLA calibration pipeline within the Common Astronomy Software Application (CASA; \citealt{CASA2022}). 

\subsubsection{6~GHz (C-band) VLA Observations}

The C-band observations span a total bandwidth of 4GHz, centered at 6GHz and consist of both A configuration (Project ID: 18A-399, PI: Kamieneski, P.; see \citealt{Kamieneski_2024}) and C configuration data sets (Project ID: 22B-078, PI: Harrington, K.). Both datasets use bandpass calibrator 3C286 (J1331+305) and amplitude calibrator J1146+3958. The A configuration dataset has a total observing time of 3 hours, split between two sources and resulting in 0.76 hours of on-source integration time. The largest angular scale of these observations is 8\farcs9. The C configuration dataset has a total observing time of 2.6 hours (1.3 hours on-source) and a largest angular scale of $240^{\prime\prime}$. 

To provide improved signal to noise in the A configuration observation, we apply Radio Frequency Interference (RFI) flagging with \textit{flagdata}. Imaging was performed using CASA version 6.6.5 with task \textit{tclean} with auto-multithresh masking. We employ multiscale cleaning to model the emission, and Briggs weighting with a robust parameter of 0.5. A pixel size of $0\farcs06$ was used to ensure adequate sampling of the synthesized beam. The resultant image, shown in Figure \ref{fig:cbnd}, has a synthesized beam with a FWHM along the major and minor axis $\theta_{maj} \times \theta_{min} = 0\farcs47 \times 0\farcs28$ with position angle $75^\circ$ and point-source sensitivity $3.0 ~\mu \rm Jy/\rm beam$ at the image center. The resultant image is $1.8^{\prime}$ in diameter.

We find it useful to also produce a lower resolution C-band A configuration image with matched beam shape and pixel scaling to the Ku-band image. We use this image for comparisons between the two bands, specifically spectral index analysis. This image employs a restoring beam with major and minor axis $\theta_{maj} \times \theta_{min} = 1\farcs45 \times 1\farcs28$ and position angle $77^\circ$. We use a pixel scale of $0\farcs25$ to adequately sample this beam. Otherwise, the imaging process is identical to the higher-resolution image. The final image has a point-source sensitivity of $5.8 ~\mu \rm Jy~ \rm beam^{-1}$ and diameter of $7.5^{\prime}$.

Initial imaging of the C configuration observation exhibited significant artifacts, mainly strong sidelobes radiating from both NATs. To address this RFI was removed with \textit{flagdata}. Then, two rounds of phase calibration using \textit{gaincal} were applied, which successfully removed most artifacts. We attempted to combine the A and C configuration observations, but the excess in short baselines resulted in significantly lower resolution than the standalone A configuration image. The lack of short baselines in the A configuration image could result in a loss of sensitivity at large spatial scales, but we have compared its total flux to the C configuration data and find the values consistent within error bars. We therefore determine the A configuration image does not resolve out any significant flux components, and we use only this image in our further analysis. 

\subsubsection{15~GHz (Ku-band) VLA Observations}
\begin{figure*}
    \centering
    \includegraphics[width=\textwidth]{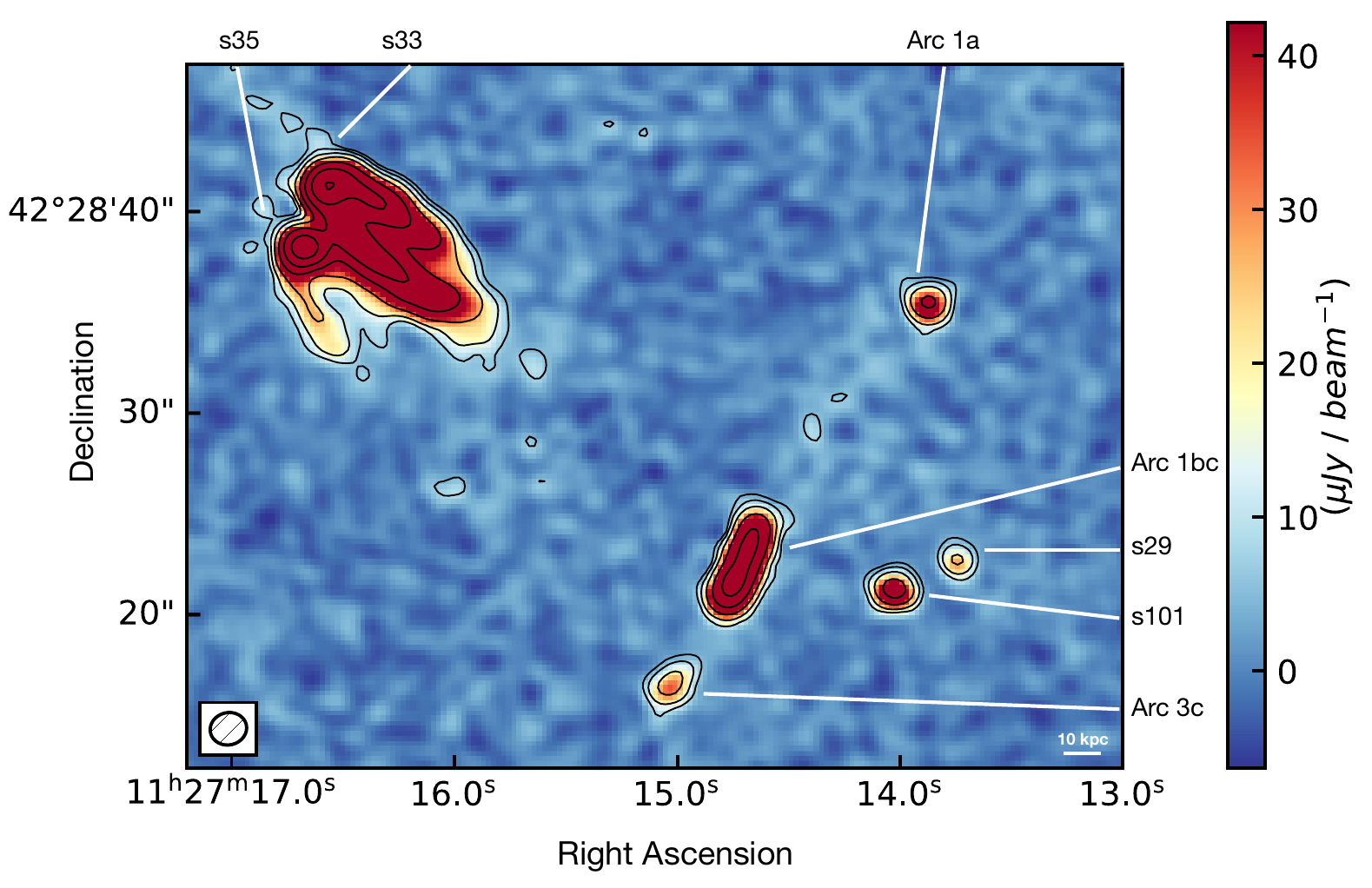}
    \caption{Ku-band VLA image of the central region of the G165 cluster with the same field of view and orientation as Figure \ref{fig:nircam}. The same set of radio sources are detected as seen in Figure \ref{fig:cbnd}. Beam is in the lower left corner. The contours shown start at $3\sigma$ and increase by factors of 2. The scale bar in the lower right assumes $z=0.348$.}
    \label{fig:kubnd}
\end{figure*}

The Ku-band observations (Project ID: 22B-078, PI: Harrington, K.) consist of 3.02 hours of on-source integration time in C configuration. The observations span a total bandwidth of 6GHz, centered at 15GHz. The largest angular scale of these observations is $97^{\prime\prime}$. This data set uses bandpass calibrator 3C286 and amplitude calibrator J1130+3815. These observations exhibited strong sidelobes radiating from s35 (brightest source in the field, see Figures \ref{fig:cbnd} and \ref{fig:kubnd} for label), similar to those seen in the C configuration C-band data. We performed RFI flagging using the same method as for C-band. Phase and amplitude calibrations were attempted, but not applied, as they either introduced additional artifacts or resulted in excessive flagging. Imaging followed the same procedure as for the C-band: auto-multithresh masking, multiscale cleaning, and Briggs weighting with a robust parameter of 0.5. A pixel size of $0\farcs25$ was used to ensure adequate sampling of the synthesized beam. This final image, seen in Figure \ref{fig:kubnd}, has a synthesized beam with a FWHM along the major and minor axis $\theta_{maj} \times \theta_{min} = 1\farcs45 \times 1\farcs28$ with position angle $77^\circ$ and point-source sensitivity $1.8 ~\mu \rm Jy/\rm beam$ at the image center. The resultant image is $4.2^{\prime}$ in diameter.

\subsection{JWST NIRCam}

A total of five epochs of JWST Near Infrared Camera (NIRCam) observations were obtained in 6-8 filters in the G165 cluster field. These data were obtained from the Mikulski Archive for Space Telescopes (MAST) at the Space Telescope Science Institute. The specific observations analyzed can be accessed via \dataset[doi:10.17909/ygd9-m030]{https://doi.org/10.17909/ygd9-m030}. The data reduction and analysis of epochs 1-3 are described in \citet{Frye_2024}. Epochs 4 and 5 were observed by JWST GO program PID 4744 (PIs Frye \& Pierel). The exposures were obtained using both modules of NIRCam in six filters: F090W, F150W, F200W, F277W, F356W, and F444W. The data reduction was carried out by a similar approach to that in epochs 1, 2, and 3, and the five epoch image was shown in \citet{Agrawal_2026}.

Figure 1 depicts the final full-depth NIRCam mosaic, from all five observing epochs combined, in the central region of G165 in the main NIRCam mosaic.

\section{Results} \label{sec: results}
\begin{deluxetable*}{ccccccccccccc}
\tabletypesize{\scriptsize} 
\label{tab:cluster mems}
\tablecaption{G165 Cluster Member Radio Properties \label{tab:results}}
\tablehead{
\colhead{ID} & \colhead{RA}& \colhead{Dec}& \colhead{$z$}&  \colhead{$F_{\rm 6 GHz}$}& \colhead{$F_{\rm 15GHz}$}& \colhead{Spatial}& \colhead{Spatial}&  \colhead{$\alpha^{\rm 6GHz}_{\rm 15GHz}$}& \colhead{$v_{\rm gal,t}$}& \colhead{$v_{\rm gal,t}$}&\\[-3mm]
\colhead{} & \colhead{}& \colhead{}& \colhead{}&  \colhead{}& \colhead{}& \colhead{Ext.}& \colhead{Ext.}&  \colhead{}& \colhead{NHFM}& \colhead{MCM}&\\
\colhead{} &\colhead{(h:m:s)}& \colhead{(d:m:s)} & \colhead{}& \colhead{($\rm mJy$)}& \colhead{($\rm mJy$)}& \colhead{$\prime\prime$}&  \colhead{($\rm kpc$)}& \colhead{}& \colhead{($\rm km ~\rm s^{-1}$)}& \colhead{($\rm km ~\rm s^{-1}$)}& \\
}
\startdata 
 s33& 11:27:16.59& 42:28:41.25& $0.348$& $5.77 \pm 0.29$& $2.90 \pm 0.16$& 12.1&  60&  $-0.76 \pm 0.04$& $2000^{+3800}_{-2000}$& $2700 \pm 1100$\\
 s35& 11:27:16.69& 42:28:38.16& $0.3375$& $1.334 \pm 0.073$& $0.623 \pm 0.063$& 6.1&  31&  $-0.78 \pm 0.04$& $2200 \pm 1500$& $1700 \pm 600$\\
 s29& 11:27:13.76& 42:28:22.58& $0.3477$& $0.065 \pm 0.013$& $0.0239 \pm 0.0070$& 0.24&  1.2&  $-0.75 \pm 0.13$&---&  ---&\\
 s101& 11:27:14.04& 42:28:21.18& $0.3427$& $0.151 \pm 0.036$& $0.076 \pm 0.019$& ---&  ---&  $-0.77 \pm 0.13$ &---&  ---&\\
.\enddata
\tablecomments{We use the same naming convention as \cite{Frye_2024, Frye_2019}. Right Ascensions and Declinations listed are the positional centroid of the higher resolution C-band imaging for each source. Redshifts are sourced from \cite{Frye_2024}. Flux densities are computed within the $3\sigma$ contour of each source, $F_{6GHz}$ corresponding to the C-band flux density and $F_{15GHz}$ corresponding to the Ku-band flux density. Spatial extent listed is the distance between furthest points of a source, detected above the $3 \sigma$ level, in the high resolution C-band image. For the NATs, this is the distance from the vertex to the end of the tails marked by the furthest pixel with SNR $> 3$. For s29 this is the FWHM of a gaussian fit for the source, deconvolved with the beam. We do not report a spatial extent for s101 because it is unresolved, and thus cannot be deconvolved with the beam. The spectral index values listed are the mean of the flux-weighted average  spectral index value in several different apertures centered on the galaxy center. Additional information about these measurements is included in Appendix \ref{sec: appendix}. Measured transverse galaxy velocities are listed for both the nonrelativistic hydrodynamic flow model (NHFM) and the Mach cone model (MCM).\\}
\end{deluxetable*}

We detected a total of six galaxies: four  in the cluster (s33, s35, s29, and s101) and two in the background (Arc 1a, Arc 1bc, and Arc 3c; $z=2.2$), where Arc 1 was strongly-lensed into two images. We use the same naming convention as in \cite{Frye_2019, Frye_2024}, where s33 and s35 refer to the North and South NAT, respectively, and s35 is the brightest cluster galaxy (BCG). Physical characterizations of all six objects, including measurements of flux, physical extent, and positional centroids, are described in Appendix \ref{sec: appendix}. Measurement values are listed in Table \ref{tab:cluster mems} (main text) for cluster member galaxies and Table \ref{tab:arcs} (Appendix \ref{sec: appendix}) for lensing arcs.

\begin{figure}
    \centering
    \includegraphics[width=0.5\textwidth]{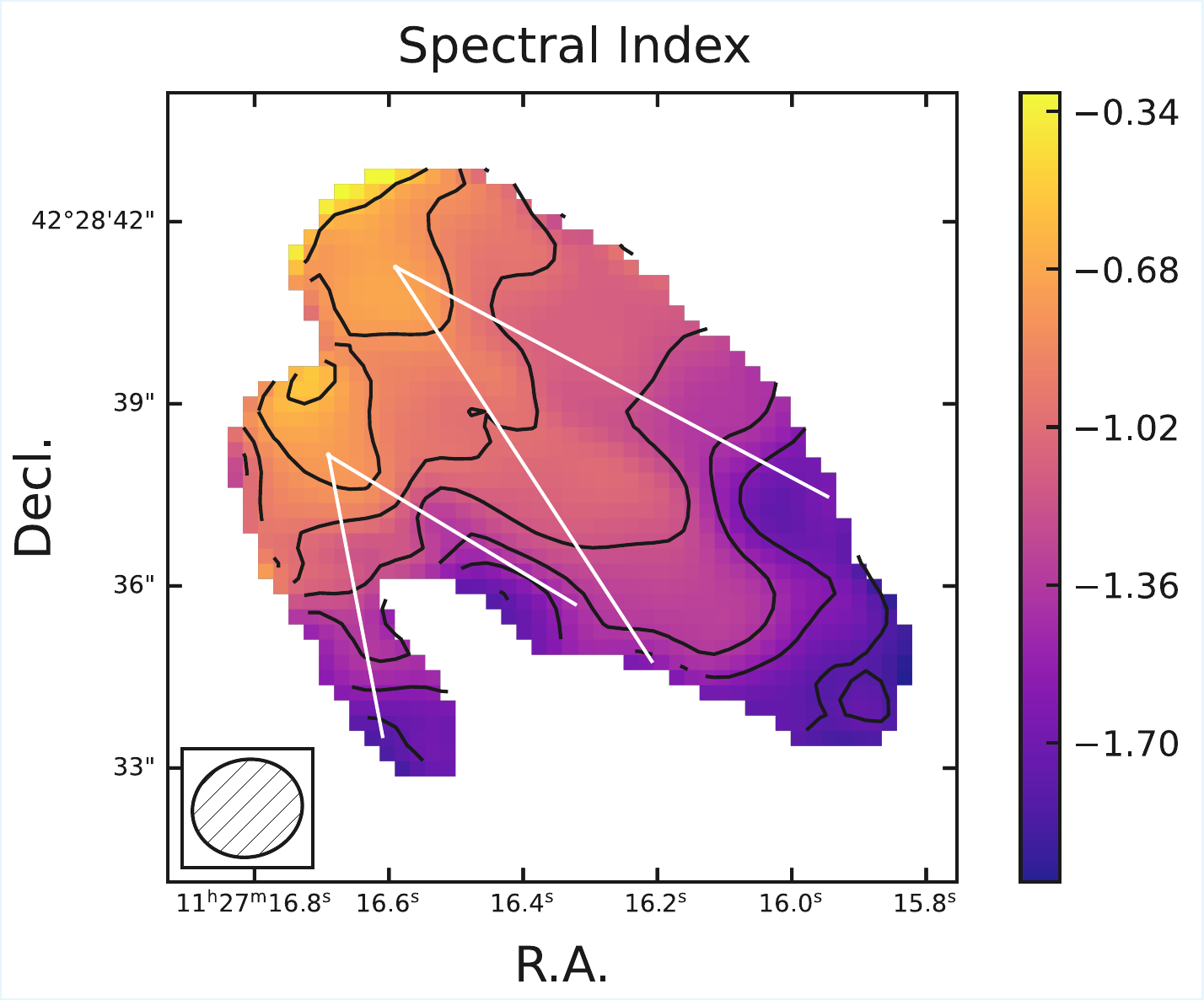}
    \caption{Spectral index map of s35 and s33 at $ \sim 1\farcs5$ resolution of C- and Ku- band, where initial images are clipped at 3$\sigma$. Beam size is in the lower left corner. The contours start at -1.8 and increase linearly in increments of 0.2. The full opening angle (see Figure \ref{fig:cartoon}) of each NAT is marked in white, with the location of the vertex corresponding to the galaxy centroid listed in Table \ref{tab:cluster mems}.}
    \label{fig:spix l}
\end{figure}

A spectral index map was made using the Ku-band image and the lower resolution C-band image discussed in Section \ref{sec: data redux}. We exclude LOFAR data for the cluster (discussed in \citet{Pascale_2022}) from this analysis due to its low resolution. Pixels with SNR $< 3$ in each image were masked and CASA task \textit{immath} in mode \textit{spix} was used to create the spectral index map (Figure \ref{fig:spix l}). Spectral index was computed as follows:

\begin{equation}
    \alpha = \frac{\ln(I_0/I_1)}{\ln(\nu_0/\nu_1)}
\end{equation}

where $I_0/I_1$ is the ratio of intensities between the C- and Ku-band images, and $\nu_0/\nu_1$ is the ratio of their central frequencies. A gradient in spectral index is found from vertex to tail in each NAT. Figure \ref{fig:spix l} shows that the spectral index values are flatter near the center of each NAT and steeper toward the tails.

Alongside the spectral index map in Figure \ref{fig:spix l}, we obtain an average spectral index value for the cluster galaxies. This is done by taking the flux-weighted mean of all pixels in the spectral index map within a circular aperture centered on the galaxy coordinates reported in Table \ref{tab:cluster mems}. The aperture sizes used are 2\farcs0, 1\farcs5, 1\farcs0, and 0\farcs5 for both NATs, 1\farcs5, 1\farcs0, and 0\farcs5 for s101, and 1\farcs0 and 0\farcs5 for s29 (omitting larger apertures for s101 and s29 due to their smaller angular sizes). Reported spectral index values are the mean of each measurement with standard deviations as the uncertainty. 

While we lack the spatial resolution to make spectral index maps of the galaxies s29 and s101 in the southwest half of the cluster (the emission is confined to an area the size of $\sim 2$ beams),  we still report the average spectral index value. The uncertainty is computed as the propagated error of the flux measurements, as we do not have the spatial resolution to significantly vary the aperture size. These values can be found in Table \ref{tab:cluster mems}.

As expected of the most massive galaxies in a cluster, s33, s101, and s29 are all situated at or near the redshift of the cluster \citep{Frye_2019}. By contrast, the BCG (s35) has a radial velocity of $v_r = -3300 ~\rm km ~\rm s^{-1}$, one piece of evidence indicating a significant disturbance \citep{Frye_2019}. The presence of radio tails informs us that there must also be a transverse velocity component typically $\gtrsim 1000 ~\rm km ~\rm s^{-1}$ \citep{Begelman_1979, ODea_1985, Bliton_1998}. Moreover, s33 and s35 both present twin tails of nearly equal amplitudes, suggesting that the NATs are not rotated about the axis of transverse motion. 
As a result, the observed bending is likely intrinsic rather than a consequence of projection effects. This configuration provides a unique opportunity to characterize the kinematics of these disturbed galaxies by using the available observational constraints to estimate the transverse velocity of both NATs through two complementary approaches: a Mach-cone model and a nonrelativistic hydrodynamic flow model, discussed in turn below.

\subsection{Transverse Velocities: Mach Cone Model} \label{subsec: mach cone}

Initial models of NATs assume a pressure equilibrium between the ICM gas and the tails, allowing the morphology to be approximated by a shock wave cone (or Mach cone) \citep{Jaffe_1973, Begelman_1979}. These early studies stopped short of estimating the transverse velocity, a step more recent works have taken \citep{Schellenberger_2017, Rajpurohit_2024}.

In this Mach cone model, a relation is obtained 
between the opening angle, galaxy velocity, and sound speed. The Mach number can be written as:
\begin{equation}
    \mathcal{M} \equiv \frac{v_{\rm gal,t}}{v_{\rm sound}} = \frac{1}{\sin{\theta}}
\end{equation}
where $\theta$ is half the opening angle of the jets.
The  opening angle, shown in Figure \ref{fig:cartoon}, can be constrained directly from our observations in two complementary ways. First, we sample the tail in bins at increasing distances from the galaxy centroid. For each bin the maximum pixel is taken from both tails and the angle subtended by these points and the galaxy centroid is measured. We use the mean of this set of angles to compute $\theta_{s33} = 15^\circ \pm 6^\circ$ and $\theta_{s35} = 24^\circ \pm 9^\circ$, where uncertainties are the standard deviations. The full opening angle ($2\theta$) is plotted in Figure \ref{fig:spix l}. A second method is used to verify these measurements, where a line is fit to the set of brightest pixels along the length of each tail and through the galaxy center. This method results in angles that are consistent with the first method within $5^\circ$. We note that the line of sight velocity component for each galaxy results in an apparent widening of the cone and larger measured values for $\theta$. We account for this by adding the radial velocity component in quadrature with our transverse velocity to compute a space velocity in Section \ref{sec:discussion}. Our $\theta$ measurements thus produce Mach values of $\mathcal{M}_{s33} = 4.0 \pm 1.6$ and $\mathcal{M}_{s35} = 2.5 \pm 0.9$, respectively.  

We use the sound speed of plasma at $10^7 ~\rm K$--- $\sim 500~\rm km~\rm s^{-1}$--- as a lower limit for this system \citep{Hu_2022}. The complex merging cluster NGC741 has a sound speed of $647~\rm km~\rm s^{-1}$ \citep{Schellenberger_2017}, and \cite{Donnert_2017} measures the sound speed of the Sausage cluster (CIZA J2242.8$+$5301; also a complicated merger) to be $851-918~\rm km~\rm s^{-1}$. We adopt a fiducial sound speed of $700 \pm 200 ~\rm km~\rm s^{-1}$ for G165 to reasonably bracket the range of possible speeds. This yields $v_{\rm gal,t} =2700 \pm 1100 ~\rm km~ \rm s^{-1}$ for s33 and $v_{\rm gal,t} = 1700 \pm 600 ~\rm km ~\rm s^{-1}$ for s35 (which are also listed in Table \ref{tab:cluster mems}). 

This model provides useful estimates of the transverse velocities, though it fails to incorporate all of the available geometry of the NAT morphology. We thus set out to provide a comparison by independently constraining the NAT velocities using a more involved model, the nonrelativistic hydrodynamic flow model.

\subsection{Transverse Velocities: Nonrelativistic Hydrodynamic Flow Model} \label{subsec: hydrodynamic model}

\begin{figure}
    \centering
    \includegraphics[width=\linewidth]{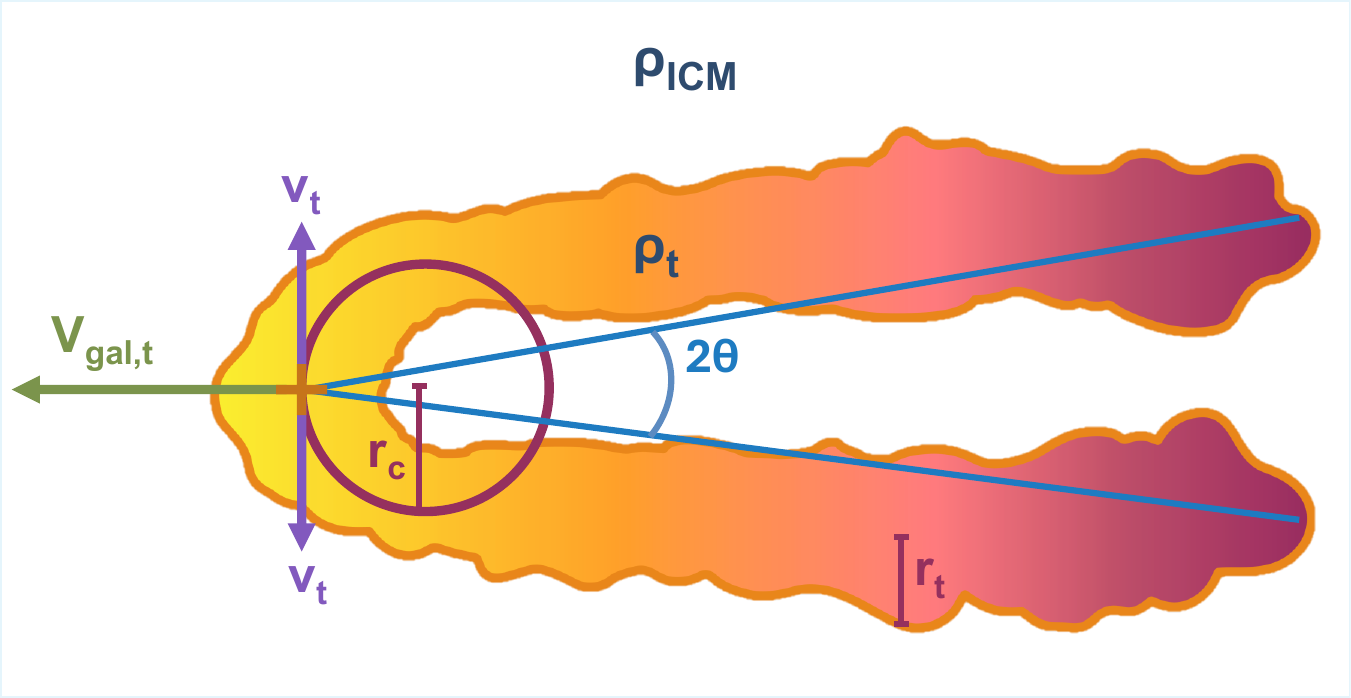}
    \caption{This cartoon shows each of the variables needed to compute the nonrelativistic hydrodynamic flow model and the Mach cone model, which are discussed in Sections \ref{subsec: hydrodynamic model} and \ref{subsec: mach cone}. We label the full opening angle $2\theta$, ICM plasma density $\rho_{ICM}$, plasma mass density of the tails $\rho_t$, cylindrical radius of the tails $r_t$, radius of curvature $r_c$, jet velocity $v_t$, and transverse galaxy velocity $v_{\rm gal,t}$. The galaxy centroid is marked with an orange cross.}
    \label{fig:cartoon}
\end{figure}

To extend this analysis to include jet physical properties, we next apply a nonrelativistic hydrodynamic flow model to constrain the radio plasma properties responsible for the observed morphology \cite[as seen in][]{Burns_1982, Venkatesan_1994, Douglass_2008, Hu_2021}. Although radio lobes are composed of a collisionless plasma, the presence of even weak magnetic fields constrains particle motions to small gyroradii, making it reasonable to assume behavior of the plasma as a fluid \citep{Burns_1982}. Consequently, following \cite{ODea_1985}, Euler's equation can be applied in the form:

\begin{equation}
    \frac{\rho_tv_t^2}{r_c} = \frac{\rho_{\rm{ICM}}v_{\rm{gal}}^2}{r_t}
\end{equation}
where $r_c$ is the curvature radius of the tails, $\rho_t$ is the plasma mass density of the tails, $v_t$ is the speed of ejection of material in the tails from the AGN, $\rho_{\rm{ICM}}$ is the plasma density of the ICM, $v_{\rm{gal}}$ is the transverse velocity of the galaxy, and $r_t$ is the cylindrical radius of the tails. We discuss each variable in turn below, including our measurements based on the high resolution C-band imaging.

\begin{figure}
    \centering
    \includegraphics[width=\linewidth]{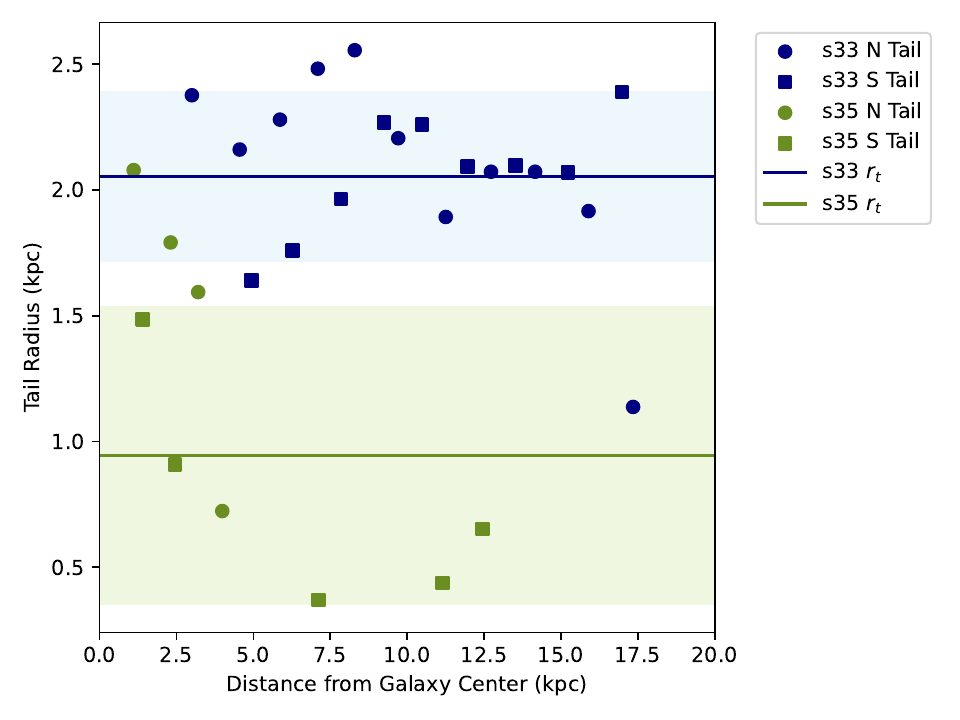}
    \caption{Measured tail radius ($r_t$) of the $3\sigma$ contour of each NAT tail. The blue and green lines are the average tail radii for s33 and s35, respectively. The shaded region around each line represents the standard deviation of the sample of measurements. The distance scaling assumes $z=0.348$.}
    \label{fig: rt}
\end{figure}

The radius of curvature of the tails--- $r_c$--- can be found by fitting a circle to the curvature of the 
jets as they are bent away from the center of the galaxy (as depicted in Figure \ref{fig:cartoon}). Operationally, we fit two circles to the emission by eye: one at the maximum and one at the minimum tolerable radius and report the mean value. We obtain $r_c = 4.60 \pm 1.02 ~\rm kpc$ for s33 and $r_c = 3.89 \pm 0.88 ~\rm kpc$ for s35, where the uncertainty captures the maximum tolerable physical extent of the jet. To determine $r_t$, we measure the width of the $3\sigma$ contour around the tails at several points along their extent, take an average, and divide by 2. We obtain $r_t = 2.06 \pm 0.34 ~\rm kpc$ for s33 and $r_t = 0.95 \pm 0.60 ~\rm kpc$ for s35. Figure~\ref{fig: rt} shows the measurements at $r_t$ at various intervals along the tails for s33 (shown in blue) and s35 (shown in green). The reported uncertainties are computed as the standard deviations of these measurements. As seen in Figure \ref{fig: rt}, there is a wide range of measured tail radii, particularly in the North tail of s35. We note that this measurement scheme for $r_t$ and $r_c$ may be impacted by the angular resolution of our observations, but we expect this effect will be much smaller than the uncertainties we adopt.

The jet velocity, $v_t$, cannot be directly inferred from the geometry, but can nonetheless be constrained from the spectral index and flux measurements. The jets in both of our sources are resolved, low brightness, and non-dominant with respect to the core emission and thus classified as FR-1 types \citep{Faharoff_1974}. These jets have a nonrelativistic (and at most mildly relativistic) jet speed and as such we can estimate $v_t$ following the prescription in \cite{Terni_de_Gregory_2017}, based on the derivation in \cite{ODea_1985}.  We define the ratio $R_{j-cj}$ of jet and counter-jet brightnesses as the ratio in flux between the brighter fainter tail. This ratio is related to the jet speed in the following manner:
\begin{equation}
    R_{j-cj} = \left(\frac{1+\beta_j\cos{\phi}}{1-\beta_j\cos{\phi}}\right)^{2+\alpha}
\end{equation}
\\
where $\beta_j = \frac{v_t}{c}$, $\alpha$ is the spectral index, and $\phi$ is the angle of motion with respect to the line of sight (which we call the inclination angle). We measure $R_{j-cj}$ by finding the flux density in each tail enclosed by the $3\sigma$ and $5\sigma$ contours and taking the flux ratio between tails. We propagate the flux density uncertainties to compute an uncertainty on each of these ratios. The ratios presented are the mean of the 3$\sigma$ and 5$\sigma$ values. The associated uncertainty is defined as the larger of the individual ratio uncertainties or the discrepancy between the two ratios. This results in $R_{j-cj}$ values of $1.09 \pm 0.05$ for s33 and $1.24 \pm 0.04$ for s35. These ratios are closer to unity than the sources in \cite{Terni_de_Gregory_2017}, further suggesting our sources are less relativistic. The radial components of the each NAT's velocities are approximately $700 ~\rm km~\rm s^{-1}$ (s33) and $3300 ~\rm km ~\rm s^{-1}$ (s35), as measured from their redshifts compared to the cluster redshift. Because of its relatively small radial velocity component relative to the high transverse component (as inferred from the Mach cone model estimate in Section \ref{subsec: mach cone} and the presence of nearly-parallel tails), we estimate s33 to have an inclination angle of $70^\circ \pm 10^\circ$. For s35, we estimate a more intermediate inclination angle of $55^\circ \pm 10^\circ$ given its very high radial velocity and pronounced, nearly parallel tails. We estimate intermediate inclination angles because they are more statistically likely. We test with larger uncertainties and find that the final galaxy velocity value and uncertainty is not significantly impacted. 
On applying our measured spectral index values from  Table \ref{tab:cluster mems}, we obtain jet speeds of $v_{t,s33} = 0.10^{+0.19}_{-0.07}~c$ and $v_{t,s35} = 0.15^{+0.09}_{-0.05}~c$. These speeds are similar to those obtained for NATs in other systems \citep{ODea_1985, Terni_de_Gregory_2017, Douglass_2008, Hu_2021}. 

To estimate $v_{\rm gal,t}$ with the nonrelativistic hydrodynamic flow model, the last quantity needed is the relative mass density of the tails and the ICM. \cite{Guo_2015} simulates AGN and defines possible ratios for $\eta \equiv \rho_{\rm t}/\rho_{\rm ICM}$ to be between 0.001 (light jet) and 0.1 (heavy jet). We thus compute galaxy velocities for $\eta = 0.001, 0.01, 0.1$. The resulting values for $\eta = 0.1$ ($v_{\rm gal,t,s33} = 6300^{+12000}_{-6300} ~\rm km ~\rm s^{-1}$ and $v_{\rm gal,t,s35} = 7000\pm 4800 ~\rm km ~\rm s^{-1}$) are unphysically large, and are excluded from further analysis \citep{Windhorst_2018}. The $\eta = 0.001$ scenario results in $v_{\rm gal,t,s33} = 630^{+1200}_{-630} ~\rm km ~\rm s^{-1}$ and $v_{\rm gal,t,s35} = 700\pm 480 ~\rm km ~\rm s^{-1}$. We favor the $\eta = 0.01$ scenario, as explained in Section \ref{subsec: kinematics}, and those velocities are listed in Table \ref{tab:cluster mems}.

\section{Discussion} \label{sec:discussion}

\subsection{NAT Kinematics} \label{subsec: kinematics}

We measured the geometric properties of the two NAT galaxies in the G165 cluster in two ways. The results of the nonrelativistic hydrodynamic flow model are sensitive to the value for the density parameter. We favor a value in the intermediate range of what is typically used of $\eta = 0.01$, which translates to transverse velocities of $v_{\rm gal,t,s33} = 2000^{+3800}_{-2000} ~\rm km ~\rm s^{-1}$ and $v_{\rm gal,t,s35} = 2200 \pm 1500 ~\rm km ~\rm s^{-1}$. This is based on estimates of typical plasma mass densities of the ICM and  tails that are rough and made independent of each other, such that constraining their ratio ($\eta$) remains highly uncertain \citep{Venkatesan_1994, Freeland_2011}. Some observational studies have found preference for the $\eta = 0.01$ scenario, because it produces velocities most likely to result in the morphology peculiar to NATs \citep{Bliton_1998, Hu_2021}. Lighter jets with lower values for $\eta$ are also consistent with simulations that show a low tail-to-ICM density ratio that allows more efficient momentum transfer from the ICM to the tails \citep{O'Neill_2019}. We note that the velocity result that assumes $\eta = 0.01$ is consistent with the result obtained by applying the Mach cone model.

The transverse velocities obtained in this work are slightly larger than the prediction of $100-1700~\rm km ~\rm s^{-1}$ in \cite{Pascale_2022}.  Other works often approximate NAT transverse velocities to be $\approx 1000 ~\rm km ~\rm s^{-1}$ \citep{Bliton_1998, Muller_2021}, based on simulations which find velocities of a minimum of $\approx 1000~\rm km ~\rm s^{-1}$ are required to produce NAT morphology. In addition, \citet{Schellenberger_2017} \& \citet{Rajpurohit_2024} use the Mach cone model to measure transverse velocity components of $\sim 1300 ~\rm km ~\rm s^{-1}$ for the NAT housed by NGC 741. Variations of the nonrelativistic hydrodynamic flow model have resulted in transverse velocity measurements of $\approx 600 ~\rm km ~\rm s^{-1}$ for NGC 1265, NGC 742, NGC 1044, NGC 4061, NGC 7503, and the NAT in Abell 1446 \citep{O'Dea_1986, Venkatesan_1994, Douglass_2008}. This method has also been used to measure $v_{\rm gal,t} \approx 2800 ~\rm km~\rm s^{-1}$ for the NAT housed by the Abell 1775 cluster \citep{Hu_2021}. A combination of these two methods results in a galaxy velocity of $920 ~\rm km ~\rm s^{-1}$ for the NAT in the 1919$+$479 cluster \citep{Burns_1986}. Very little comparison has been done between the two different models, but we note that the Mach cone model produces smaller error bars. This is likely due to the smaller number of input parameters, and may not reflect a more accurate measurement. There does not appear to be a systematic offset between the results of these models in the literature, nor do we observe one here. We recognize that our choice of model may significantly impact the results of our study, hence our choice to employ two independent models. We select the Mach cone model for its simplicity and the nonhydrodynamic flow model for its many observable constraints. We take the rough agreement of the models as evidence that, despite large uncertainties, these models are useful for describing the morphology and dynamics of the NATs in G165.

To further contextualize our result, we use the measured radial velocity of the brightest cluster galaxy (BCG; s35) of $|v_r| \approx 3300 ~\rm km ~\rm s^{-1}$ \citep{Frye_2024}, added in quadrature to the transverse velocity components, to calculate a total space velocity. The nonrelativistic hydrodynamic flow model and the Mach cone model result in space velocities of $v_{\rm space} \approx 4000 ~\rm km ~\rm s^{-1}$ and $v_{\rm space} \approx 3700 ~\rm km ~\rm s^{-1}$, respectively. In some cases, velocity dispersion has been used as a substitute for galaxy space velocity in NAT computations, where $v_{\rm space} \sim \sqrt{3}\sigma$ \citep{Burns_1986, Banfield_2016}. This approximation is in agreement with our measurement, as $\sigma = 2400 ~\rm km ~\rm s^{-1}$ \citep{Frye_2019} results in $v_{\rm space} \approx 4100 ~\rm km ~\rm s^{-1}$. We suggest that the high velocity relative to the cluster mean, and the highly disturbed state of the cluster (and thus the large velocity dispersion) may explain our large space velocities.

\subsection{Spectral Index Analysis} \label{subsec: shocks}

The spectral index map derived from our 6GHz and 15 GHz VLA observations (Figure \ref{fig:spix l}) shows a clear steepening gradient along both NAT tails. The spectral index is relatively flat ($\alpha \sim -0.3$) near the head and steepens to $\alpha \sim -1.8$ at the ends of the tails. This result is consistent with expectations of a central engine injecting energetic electrons (flatter $\alpha$) that lose energy as they propagate through the ICM 
\citep{Jaffe_1973, Feretti_1999, Rajpurohit_2024, Gani_2026}.
As the energetic electrons diffuse, the tail is primarily composed of aging electrons resulting in a steeper $\alpha$. Magnetohydrodynamic simulations of NAT formation predict similar gradients for NATs at this stage in their formation, and suggest that a flatter spectral index is related to larger magnetic fields towards the galaxy centers \citep{O'Neill_2019}. 

The spectral index gradient is relatively smooth and shows no significant flattening or re-brightening along the tails. The absence of regions of  re-flattened spectral index suggests merger shocks in the cluster either have not yet formed or have not yet intersected the NAT tail regions. 
The high galaxy velocities and the apparent dynamical youth of the cluster suggests that the NAT tails formed near to or even during the galaxies' initial in-fall, and may still be on their first approach to the dense, shock-active regions of the ICM. X-ray observations, both in hand and approved, will be essential for identifying the thermodynamic state of the intracluster gas and any morphological clues that could test this hypothesis and further constrain the merger timeline (XMM AO22 Proposal ID 92030, PI B. Frye, and an in-execution joint Chandra/XMM program Chandra GO 27620212, PI M. Donahue.).

\subsection{Dynamical State} \label{subsec: dynamics}

Both NAT galaxies exhibit nearly parallel tail morphologies extending toward the southwest, away from the cluster center. Simulations predict that newly formed NAT tails undergo a ``flapping" motion during their first few Myr, producing sharp bends that become less pronounced as the system matures \citep{O'Neill_2019}. This intermediate phase of head-tail formation produces a distinct `$\Omega$' tail morphology in observations of the cluster J0321$-$455—housing both a NAT and its lower-velocity counterpart, a Wide-Angle Tail (WAT) galaxy—and the dual WATs in Abell 400 \citep{Owen_1985, Klamer_2004}. \cite{Bruno_2024} reports wiggles in the tails of the two NATs in Abell 2142 also indicative of this formation process. The absence of such features in our observations suggests NAT lifetimes $\gtrsim 100 ~\rm Myr$ \citep{O'Neill_2019}. This is similar to the timescale estimated for the NAT sources in the seven clusters discussed in \citet{Sebastian_2017} as well as in the IIZW108 cluster \citep{Bruno_2024}, which exhibit similarly mature morphology to the NATs in G165. This timescale, combined with the galaxies' apparent motion away from the cluster core, supports a scenario in which both systems are positioned near to the center of the cluster.

The dynamical state of the cluster is further constrained by the large space velocities we measure (see Section \ref{subsec: kinematics}), which approach the theoretical maximum in-falling velocity for relaxed galaxy clusters \citep{Windhorst_2018} and the highest allowed velocity expected for G165, $v_{\rm space} \sim \sqrt{3}\sigma$ \citep{Frye_2019}. These velocities indicate that the BCG (s35) is not yet kinematically relaxed within the global potential of G165, but rather is involved in an active ongoing major cluster interaction. This result is in agreement with the interpretation of the cluster dynamics found in \citet{Pascale_2022}. In addition to our support of this disturbed cluster environment scenario, our result indicates that s35 is situated spatially at or near the base of the gravitational potential well, where the maximum theoretical infall velocity occurs. 

G165 presents a rare case of two NAT galaxies with high velocities that are in the same cluster, a configuration seen in only a handful of other systems. Similar clusters housing two NATs (or WATs, in some cases) show that NATs are typically found in-falling into the cluster's gravitational potential well \citep{Jaegers_1983, Downes_1984, Muller_2021, Bruno_2024} or in a gravitationally bound binary with another head-tail galaxy \citep{Klamer_2004}, while WAT galaxies are more commonly found in the cluster center \citep{Jaegers_1983, Owen_1985}. Here we observe two NATs moving in the same direction, and away from the cluster center. This suggests a post-pericentric cluster orientation, where the NAT morphology is a result of sub-cluster motion away from the cluster center of mass.

In this scenario, we expect the X-ray peak to be offset from either of the two mass peaks. Indeed, the first XMM observations of G165 (M. Donahue and B. Frye, private communication) show a clearly-detected and extended X-ray source centered on (R.A., Dec.) = (11:27:15.52, +42:28:27.0) and situated between the two mass peaks. The new observations from Chandra will yield high angular resolution that will be far more sensitive to the distribution of X-rays across the cluster. The second epoch of XMM observations will yield improved temperature and abundance constraints, as well as improved signal to noise outside the main core region. In an upcoming paper, we will provide a full description of all of these observations and their analysis. 



\section{Conclusion} \label{sec: conc}

In this work, we present the first 15GHz VLA observations and a new, higher-resolution reduction of 6GHz observations in the G165 cluster field. The cluster houses two Narrow Angle Tail galaxies which both exhibit nearly parallel tails extending toward the southwest, indicating motion through the cluster and away from the cluster center of mass. We make the first observationally-based measurements of the transverse velocities of these galaxies using two independent models. This analysis results in transverse velocities for both NATs of $\sim 2000 ~ \rm km ~\rm s^{-1}$, resulting in a space velocity of $\sim 4000 ~\rm km ~\rm s^{-1}$ for the BCG (s35). This extreme velocity, approaching the theoretical free-fall limit for cluster members \citep{Windhorst_2018, Frye_2019}, suggests the BCG is currently transiting the cluster core, where it is subjected to maximum ram pressure and merger-driven shocks. In addition to this dynamic analysis, we produce the first spectral index map of this region, providing constraints on electron aging and cluster shocks. The absence of re-acceleration features in both NATs suggests an early stage cluster interaction where merger-driven shocks have not yet had time to interact with the NAT tails.

Further work on the dynamics of this cluster will be discussed in a future publication, which will discuss X-ray observations. These observations will help constrain the presence and orientation of cluster shocks as well as ISM stripping of the NAT galaxies. Additional context for the dynamics of the cluster will be obtained by analysis of radio polarization observations (Project ID: 22B-078, PI: Harrington, K.), which will provide magnetic field constraints on NAT tail aging and bending.

\begin{acknowledgements}
AW acknowledges support from the University of Arizona W.A. Franke Honors College Summer 2025 Stipend 000006. AW thanks Jamie Wilde for her assistance in designing Figure \ref{fig:cartoon}. AW also acknowledges Rachel Honor and Nikhil Garuda for their support in group meetings. 

BLF acknowledges support from JWST programs GO-4446 and GO-4744.

BLF and MD also acknowledges support from XMM, Proposal ID 92030.

RAW acknowledges support from NASA JWST Interdisciplinary Scientist grants NAG5-12460, NNX14AN10G and 80NSSC18K0200 from GSFC.

PSK acknowledges financial support from the Knut and Alice Wallenberg Foundation.

The National Radio Astronomy Observatory and Green Bank Observatory are facilities of the U.S. National Science Foundation operated under cooperative agreement by Associated Universities, Inc.

\end{acknowledgements}
\facilities{VLA, JWST}
\software{Astropy \citep{Astropy_2013, Astropy_2018, Astropy_2022}, CASA \citep{casa}, Matplotlib \citep{matplotlib}}

\appendix
\section{Radio Imaging Details, Source Positions, and Flux Densities} \label{sec: appendix}
We note the marginal detection of Arc 3ab (the multiply imaged counterpart of Arc 3c) in the natural-weighted reduction of these C-band observations, presented in \citet{Kamieneski_2024}, but do not detect this source in our robust 0.5 image. Radio coordinates for each cluster galaxy are found by computing a positional centroid in a $4 \times 4$ pixel region centered by eye on the vertex of the galaxy. These positions are consistent within 0\farcs5 with the brightest pixel, and are listed in Table \ref{tab:cluster mems}. We report the spatial extent for each radio source, which is defined as the extent of the $3\sigma$ contour of the high resolution C-band image. In practice, this is the diameter for circular sources and vertex-to-tail distance for the NATs. We opt for this characterization given the irregular structures of the lensed arcs and NATs.  

Flux densities were obtained for each source from both the C-band and Ku-band imaging. The reported flux densities are from the region enclosed by the $3\sigma$ contour. The uncertainties listed are the image rms noise multiplied by the square root of the region area divided by the beam solid angle. To this uncertainty we add, in quadrature, a 5\% uncertainty due to the absolute flux calibration uncertainties.  We note that Arcs 1bc and 1a may also include emission from NS\_46, which is very near in both redshift and apparent position to Arc 1. It is possible that NS\_46 is composed of tidally stripped gas from the galaxy in Arc 1 due to its interaction with a nearby galaxy, or an outflow from the galaxy in Arc 1. Further discussion of this source and its interaction with Arc 1 can be found in \cite{Frye_2024}. We refer to our detections as Arcs 1bc and 1a, with the understanding that it is possible there is an indistinguishable component of the flux from NS\_46. Our measured fluxes for Arcs 1bc, 1a, and 3c are lower than those reported by \citet{Kamieneski_2024}, likely due to resolving out some structure in our higher-resolution image. To verify this, we repeated our photometry using the lower resolution natural-weighted image from their study; this approach yielded fluxes consistent with their published values.

\begin{figure}
    \centering
    \includegraphics[width=0.5\linewidth]{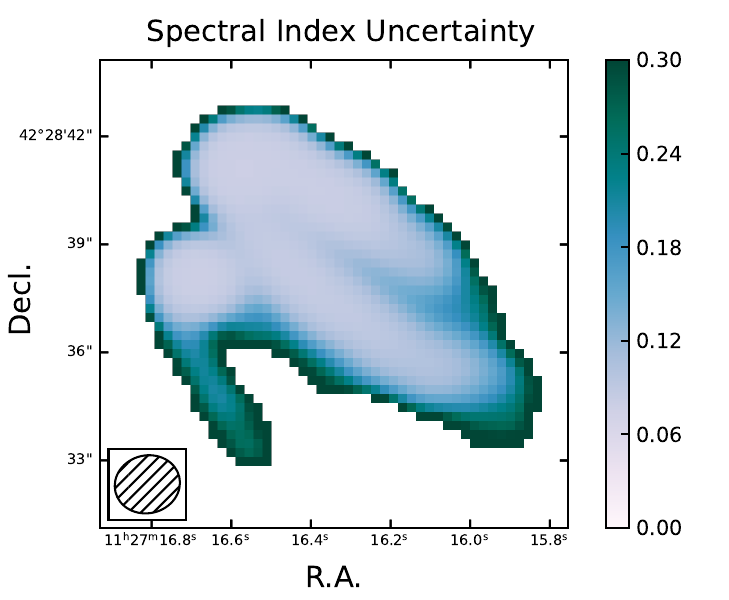}
    \caption{The spectral index uncertainty plot corresponding to the spectral index map shown in Figure \ref{fig:spix l}. Uncertainties are computed by propagating through the C-band and Ku-band flux uncertainties. Each band's uncertainty is the image RMS added in quadrature to a 5\% total absolute flux calibration uncertainty. Beam size is in the lower left corner.}
    \label{fig:spix uncert}
\end{figure}

In addition to the spectral index map (Figure \ref{fig:spix l}), we produce a spectral index uncertainty map, shown in Figure \ref{fig:spix uncert}. To do this, we propagate the errors in flux from each VLA band through to compute a spectral index uncertainty for each pixel, as given by the following relation:

\begin{equation}
\sigma_{\alpha} = \frac{1}{\ln(\nu_0/\nu_1)} \sqrt{\left(\frac{\sigma_{I_0}}{I_0}\right)^2 + \left(\frac{\sigma_{I_1}}{I_1}\right)^2}
\end{equation}

We define the flux uncertainties ($\sigma_{I_0}, \sigma_{I_1}$) as the RMS of the image and a 5\% absolute flux calibration error, added in quadrature. The uncertainty map that we produce shows higher errors near the edges of the NATs, where there is less emission. 

\begin{deluxetable*}{cccccccccc}
\tabletypesize{\scriptsize} 
\label{tab:arcs}
\tablecaption{G165 Lensing Arc Radio Properties}
\tablehead{
\colhead{Object} & \colhead{RA}& \colhead{Dec}& \colhead{$z$}& \colhead{$\mu$}& \colhead{$F_{\rm 6 GHz}$}& \colhead{$F_{\rm 15GHz}$}& \colhead{Spatial Extent}&  \colhead{$d_{\mu}$}&\\
\colhead{} &\colhead{(h:m:s)}& \colhead{(d:m:s)} & \colhead{}& \colhead{}& \colhead{($\rm \mu Jy$)}& \colhead{($\rm \mu Jy$)}& \colhead{$\prime\prime$}&  \colhead{($\rm kpc$)}& \\
}
\startdata 
 Arc 1bc& 11:27:14.75& 42:28:21.49& $2.2355$& $46$&  $847 \pm 46$& $443 \pm 45$& 5.2& 2.9&\\
 Arc 3c& 11:27:15.06& 42:28:16.25& $2.23$& 5.9 & $91 \pm 9$& $49 \pm 8$& 1.7& 7.3 & \\
 Arc 1a& 11:27:13.89& 42:28:35.30& $2.2355$& $6.6$& $118 \pm 13$& $70 \pm 15$& 1.27& 4.9& \\
.\enddata
\tablecomments{Right Ascensions and Declinations listed are the brightest pixel of the higher resolution C-band imaging for each source. Redshifts are sourced from \cite{Frye_2024} and magnifications are sourced from \cite{Kamieneski_2024}. Flux densities are all flux contained within the $3\sigma$ contour of each source, where $F_{\mu 6GHz}$ corresponds to the C-band flux density and $F_{\mu 15GHz}$ corresponds to the Ku-band flux density. Spatial extent listed is the distance between furthest points of a source, detected above the $3 \sigma$ level, in the high resolution C-band image. For the lensing arcs this is the length of the arcs, measured long-ways. $d_{\mu}$ is spatial extent, corrected for magnification, and converted to kpc. We assume an East-West magnification component of 3, and thus apply a magnification correction of $\frac{\mu}{3}$ when computing $d_{\mu}$.}
\end{deluxetable*}

\bibliography{ref}{}
\bibliographystyle{aasjournal}

\end{document}